\documentstyle[preprint,aps,prb,12pt]{revtex}
\def\nm{\nonumber}

\def\la{\Lambda_1}
\def\lla{\Lambda_2}
\def\llla{\Lambda_3}
\def\Del{\Delta}
\def\DDel{\Delta_1}
\def\DDDel{\Delta_2}
\def\m{m_1}
\def\mm{m_2}
\def\mmm{m_3}

\def\beqa{\begin{eqnarray}}
\def\beq{\begin{equation}}
\def\F{\cal{F}}
\def\eeqa{\end{eqnarray}}
\def\eeq{\end{equation}}
\def\lab{\label}

\begin{document}

\begin{titlepage}
%\vspace{2.0cm}
\begin{center}
%\begin{tabular}{|c|}\hline
%\\
{\Large \bf
Prepotentials}\\
\end{center}
\begin{center}
%\\
{\Large \bf of}\\ 
\end{center}
\begin{center}
%\\
{\Large \bf 
$N=2$ $SU(2)$ Yang-Mills Theories }\\
%\\
\end{center}
\begin{center}
{\Large \bf 
Coupled with Massive Matter Multiplets}\\
%\\
%\hline
%\end{tabular}
\end{center}
%\lineskip .75em
%\vskip 3em
%\normalsize
\begin{center}
{\large Y\H uji Ohta}\\
%\vskip 1.5em
Department of Mathematics\\
Faculty of Science\\
Hiroshima University\\
Higashi-Hiroshima 739, Japan.
%\vskip 1.0em
\end{center}
%\vskip2em

\begin{abstract}
We discuss $N=2$ $SU(2)$ Yang-Mills gauge theories 
coupled with $N_f \ (=2,3)$ massive hypermultiplets in the weak coupling 
limit. We determine the 
exact massive prepotentials and the monodromy matrices 
around the weak coupling limit. We also study that the double scaling limit of these 
massive theories and find that the massive $N_f -1$ theory can be obtained 
from the massive $N_f$ theory. New formulae for the massive 
prepotentials and the monodromy matrices are proposed. In these formulae, 
$N_f$ dependences are clarified.  
\end{abstract}
%\pacs{11.15, 12.60.J} 
%\begin{center}
%to be published in ``Journal of Mathematical Physics''.
%\end{center}
\end{titlepage}

\pagestyle{myheadings}
\markboth{Prepotentials of $SU(2)$ 
Yang-Mills with matters}{Prepotentials of $SU(2)$ Yang-Mills with matters}

\begin{center}
\section{Introduction}
\end{center}
\renewcommand{\theequation}{1.\arabic{equation}}\setcounter{equation}{0}

Non-perturbative properties of four dimensional $N=2$ 
supersymmetric $SU(2)$ Yang-Mills gauge theory was discussed by Seiberg and 
Witten \cite{SW1,SW2}. One of the important discoveries in 
their investigations was the fact that the quantum 
moduli space of the $N=2$ $SU(2)$ Yang-Mills theory coupled with or without 
$N_f$ hypermultiplets could be identified with 
the moduli spaces of certain elliptic curves which controlled the low energy properties. 
They could determine the exact expressions for 
the monopole and dyon spectrum and the metric on the quantum moduli space. 
Their approach was extended 
to, for example, the other gauge theory with or without matters \cite{AF,APS,KLT,D,B,HO}. In 
Ref.9, the quantum moduli space of $N=2$ $SU(2)$ Yang-Mills theories coupled with mass-less 
hypermultiplets was studied, but our knowledges for the massive 
theories are poor in contrast with the case of the mass-less theories. 
Of course Seiberg and Witten qualitatively discussed 
these massive theories in Ref.2, but we can not say that we have enough quantitative 
understandings for these massive theories because the quantitative analyses on them 
does not ever been sufficiently carried out.    

For this reason, we discussed the simplest massive theory, i.e, $N_f =1$ theory, 
in the weak coupling limit as an instructive example \cite{O}. In Ref.10, we did not 
discuss the other asymptotic free theories, i.e., $N_f =2$ and 3 because there were 
several technical obstacles in the computations 
of the periods. However, for the massive $N_f =1$, 
we could find the exact prepotential and the monodromy matrix by using Picard-Fuchs equation \cite{O}. 
We observed that the Picard-Fuchs equation was a third order differential equation. Its 
solutions could not be expressed by a hypergeometric function in contrast with the mass-less theory 
but they gave interesting informations for the massive $N_f =1$ theory. For example, the $N_f =0$ theory 
can be regarded as a low energy version of the 
massive $N_f =1$ theory \cite{SW2} and it must be obtained from the massive $N_f =1$ theory in the 
double scaling limit, but we could explicitly show how the massive $N_f =1$ theory flowed to 
the $N_f =0$ theory by using those solutions. 
On the other hand, since the massive 
$N_f =1$ theory can be regarded as a low energy theory of 
the massive $N_f =2$ theory, we are sure that the results of Ref.10 can be obtained from the 
massive $N_f =2$ theory in the double scaling limit. In general, the massive $N_f -1$ theory 
can be considered as a low energy theory of the massive $N_f$ theory. Therefore in order to 
determine the general structure of these massive gauge theories, we must extend the results of 
Ref.10. For these reasons, 
we discuss the massive $N_f =2$ and 3 theories in the weak coupling limit 
in this paper. In the text, 
we mainly discuss the massive $N_f =2$ theory because all mathematical expressions 
in the massive $N_f =3$ theory are lengthy. 
Therefore, we summarize the results for the massive $N_f =3$ theory 
in appendix C. 

The paper consists of the following sections. 
In section 2, we derive the Picard-Fuchs equation of the massive $N_f =2$ 
theory and its solutions in the weak coupling limit. 
The order of the differential equation is three as well as that of the massive $N_f =1$ equation. We 
discuss the monodromy around the weak coupling limit in the end of this 
section. The monodromy matrix can be arranged to $3 \times 3$ matrix due to the order of the 
differential equation. 
In addition to this, we will see that the monodromy matrix should be 
quantized by the winding numbers for the residues. We derive the prepotential 
and instanton contributions for it in section 3. 
The double scaling limit of this theory 
is discussed in section 4. We can find that the 
instanton expansion coefficients of the prepotential are completely coincide 
with that of Ref.10 in the double scaling limit. Final section 5 
is summary. We summarize our results as some useful formulae for $N=2$ $SU(2)$ Yang-Mills gauge theories 
weakly coupled with $N_f$ $(N_f =0,\cdots, 3)$ matter multiplets. In particular, 
general formulae of the two periods of the meromorphic 1-form, the monodromy matrices 
and the exact prepotentials are proposed. 
We emphasis that $N_f$ dependences are clarified by these results.

\section{\protect\centering $N_f =2$ Picard-Fuchs equation}
\renewcommand{\theequation}{2.\arabic{equation}}\setcounter{equation}{0}

Quantum moduli space of the $N=2$ $SU(2)$ 
Yang-Mills theory coupled with two massive hypermultiplets 
can be described by the following hyperelliptic curve
	\beq
	y^2 =\left(x^2 -u+\frac{\lla^2}{8}\right)^2 -\lla ^2 (x+\m)(x+\mm)
	\eeq
and the meromorphic 1-form \cite{HO}
	\beq
	\lambda_2 =\frac{\sqrt{2}xdx}
	{4\pi iy}\left[\frac{(x^2 -u+\lla^2 /8)(2x+\m+\mm)}
	{2(x+\m)(x+\mm)}-2x \right],
	\eeq
where $u$ is the gauge invariant parameter, 
$\m$ and $\mm$ are masses of the hypermultiplets and $\lla$ is a dynamically 
generated mass scale of the theory. 
We can also describe the same $N_f =2$ theory by an elliptic curve \cite{SW2}. 

This curve has four branching points. In the weak coupling limit $(u\longrightarrow \infty)$, 
they will behave as 
	\beqa
	x_1 &=& -\frac{\lla}{2} -\sqrt{u} +\frac{1}{\sqrt{u}}\left[ -\frac{\lla^2}{16}
		+\frac{\lla}{4}(\m+\mm)\right] +\frac{\lla}{16u}(-\m +\mm)^2 +\cdots,\nm \\
	x_2 &=& -\frac{\lla}{2} +\sqrt{u} +\frac{1}{\sqrt{u}}\left[ \frac{\lla^2}{16}
		-\frac{\lla}{4}(\m+\mm)\right] +\frac{\lla}{16u}(-\m +\mm)^2 +\cdots, \nm \\	
	x_3 &=& \frac{\lla}{2} -\sqrt{u} +\frac{1}{\sqrt{u}}\left[ -\frac{\lla^2}{16}
		-\frac{\lla}{4}(\m+\mm)\right] -\frac{\lla}{16u}(-\m +\mm)^2 +\cdots ,\nm \\
	 x_4 &=& \frac{\lla}{2} +\sqrt{u} +\frac{1}{\sqrt{u}}\left[ \frac{\lla^2}{16}
		+\frac{\lla}{4}(\m+\mm)\right] -\frac{\lla}{16u}(-\m +\mm)^2 +\cdots 
	\lab{branch}
	.\eeqa
We can take the cuts to run counter clockwise 
from $x_3$ to $x_1$ as $\alpha$-cycle 
and from $x_3$ to $x_4$ as $\beta$-cycle. The intersection number is 
$\alpha \cap \beta =1$. We can regard the curve as a genus one 
Riemann surface. 

The period integrals of $\lambda_2$ are defined by
	\beqa
	a_2 (u) &=& \oint_{\alpha} \lambda_2 , \lab{a}  \\
	a_{D}^2 (u) &=& \oint_{\beta} \lambda_2 .\lab{aD}
	\eeqa
$a_2 (u)$ is identified with the scalar component of the $N=1$ chiral multiplet and 
$a_{D}^2 (u)$ is its dual \cite{SW1,SW2}. In the mass-less theory, the above 
$\alpha$ and $\beta$ constitute canonical homology bases of the torus, but in the 
massive theory they should be replaced 
with loops so as to enclose the ``extra'' poles corresponding to $x=-\m,-\mm$. 
We will often use $\Pi_2 =\oint_{\gamma} \lambda_2$, where $\gamma$ is 
a suitable 1-cycle on the curve, as a representative of the periods. 

It is more convenient to study the Picard-Fuchs equation than the period integrals themselves 
in order to see the behaviour near $u=\infty$. The massive 
$N_f =2$ Picard-Fuchs equation is a third order differential equation which is given by
	\beq
	\frac{d^3 \Pi_2}{du^3} +\left(\frac{\Del_{2}'}{\Del_2 }
	-\frac{8D_2}{B_2}\right)\frac{d^2\Pi_2}{du^2}
	-\frac{16}{\DDDel }\left(C_2  -\frac{A_2 D_2}{B_2}\right)\frac{d\Pi_2}{du} =0,
	\lab{pic} \eeq
where 
	\beqa
	A_2 &=&-\lla^6-8\lla^4 \left[3(\m^2 +\mm^2)-u-\m\mm \right] +256u^2(\m^2+\mm^2-2u) \nm \\
	& &+32\lla^2\left[2u^2+4\m^2\mm^2 +14 \m \mm u -(\m^2+\mm^2)(3u+6\m\mm)\right],\nm \\
	B_2 &=&8\lla^4-64\lla^2\m\mm+256\left[3u(\m^2+\mm^2)-
	2u^2 -4\m^2\mm^2\right],\nm \\
	C_2 &=&\lla^4-4\lla^2\left[3(\m^2+\mm^2)-4u-12\m\mm\right]\nm \\
		& &+32\left[3u(\m^2+\mm^2)-6u^2-2\m^2\mm^2 \right],\nm \\
	D_2 &=&32\left[3(\m^2 +\mm^2)-4u\right],\nm \\
	\DDDel &=& \lla^8-48\lla^6\m\mm-16\lla^4\left[27(\m^4+\mm^4)-36u(\m^2+\mm^2)
	\right.\nm \\
	& & \left. +8u^2+6\m^2\mm^2\right] +512\m\mm\lla^2\left[9u(\m^2+\mm^2)-10u^2-8\m^2\mm^2
	\right]\nm \\
	& &+4096u^2(\m^2-u)(\mm^2-u) \lab{coepic},
	\eeqa
and $\DDDel' =d \DDDel /du$. $\DDDel $ coincides with the 
discriminant of the curve. (\ref{pic}) can be obtained from the massive $N_f =3$ 
Picard-Fuchs equation in the double scaling limit. See appendix C. 
Note that (\ref{pic}) does not show any symmetry over the $u$-plane. 
From (\ref{pic}) and 
(\ref{coepic}), we can easily find that this differential equation has regular singular points 
corresponding to $\DDDel =0$ and $B_2 =0$. We are also interested in the property near 
the discriminant loci which correspond to extra mass-less states, but the 
calculations near such singularities are not so easy, although 
they should be discussed elsewhere. 

When both of the hypermultiplets have zero mass, (\ref{pic}) 
exactly reduces to 
	\beq
	(\lla^4 -64u^2)\frac{d^2\Pi_{2}}{du^2}-16\Pi_2 =0 \lab{N2red}
	.\eeq
We set the integration constant to zero because (\ref{N2red}) can be directly 
obtained using mass-less meromorphic 1-form. The global $\mbox{\boldmath$Z$}_2$ 
symmetry over the $u$-plane is now recovered. This symmetry can appear when and only when 
both of the matters are mass-less. Therefore we can find that the masses play a role 
to break the global symmetry. (\ref{N2red}) was studied in Ref.9. 

The reader may notice that the order of (\ref{pic}) is three whereas that of 
(\ref{N2red}) is two. The mathematical background of this 
fact was discussed in Ref.10 in the case of the massive $N_f =1$ theory, so we 
briefly state here only the essence. First, recall that the massive 
meromorphic 1-form has extra simple poles corresponding to $x=-\m$ and $-\mm$. 
Since an operation of differentiating over $u$ and integrating over $x$ 
reduces the order of poles by one, the reduction will require 
one step more than the mass-less case when $\lambda_2$ is massive. Therefore the order of (\ref{pic}) and 
(\ref{N2red}) differs by one. For more complete treatment, see Refs.10 and 11. 

Next, let us try to calculate the solutions to (\ref{pic}). In order to accomplish it in the 
weak coupling limit, we introduce $z=1/u$. We find that the 
solutions to the indicial equation for (\ref{pic}) are $0,-1/2,-1/2$ (double roots). 
The solution $\rho_0 (z)$ corresponding to the index 0 
is some constant $\epsilon_2$ which may depend on $\lla,\m$ and $\mm$,
	\beq
	\rho_0 (z) =\epsilon_2 .\lab{ep}
	\eeq
At first sight, this constant solution is trivial but it is important and has non-trivial 
meanings in the massive theory, as was explicitly shown in Ref.10. 
In fact, this corresponds to 
the residue contributions of $\lambda_2$ and we can rewrite (\ref{ep}) as 
	\beq
	\rho_0 (z) =\mbox{linear combination of}\  \nu_1 \ 
	\mbox{and}\ \nu_2 ,
	\eeq 
where $\nu_1$ and $\nu_2$ are residues of $\lambda_2$. 
These constants will be determined in the comparison of the lower order expansion of the 
period integrals with fundamental solutions to the Picard-Fuchs equation. 
On the other hand, there are two independent solutions 
corresponding to the index $-1/2$. One of them is  
	\beq
	\rho_1 (z) = z^{-1/2} \sum_{i=0}^{\infty}a_{2,i} z^i ,\lab{rho1}
	\eeq
where the first several expansion coefficients $a_{2,i}$ are given 
in appendix A. We find that $a_{2,n}$ can be represented by a polynomial 
of $\lla^{2i} \m^j \mm^k$ with $2n=2i+j+k$, 
where $i,j,$ and $k$ are non-negative integers. 
$\rho_1$ coincides with a hypergeometric function in the mass-less limit \cite{IY}. 
The other solution behaves logarithmic. It is
	\beq
	\rho_2 (z) = \rho_1 (z) \ln z +z^{-1/2} \sum_{i=1}^{\infty}b_{2,i} z^i \lab{rho2}
	\eeq
where the first several coefficients $b_{2,i}$ are given in appendix B. 
$b_{2,n}$ can also be represented by a polynomial of $\lla^{2i} \m^j \mm^k$ with 
$2n=2i+j+k$ as well. Note that these 
polynomials are actually homogeneous and have obvious $\mbox{\boldmath$Z$}_2$ 
symmetry, i.e, invariance under $\m \leftrightarrow \mm$.  

We can express the periods (\ref{a}) and (\ref{aD}) 
as a linear combination of $\rho_0 ,\rho_1 $ and $\rho_2$
 by comparison with the lower order expansion 
of the period integrals. The results will be  
	\beqa
	a_2 (u) &=& \frac{\rho_1(z)}{\sqrt{2}} +n_1 \nu_1 +n_2 \nu_2 
	 , \lab{a2}  \\
	a_{D}^2 (u) &=& A\rho_2 (z) +B\rho_1 (z) +n_{1} ' \nu_1 +n_{2} ' 
	\nu_2 \lab{aD2} ,
	\eeqa
where 
	\beqa
	A&=& -\frac{i\sqrt{2}}{2\pi} ,\nm \\
	B&=& \frac{i\sqrt{2}}{2\pi}(-2+4\ln 2 +\pi i -2\ln \lla ),\nm \\
	\nu_1 &=&- \frac{\sqrt{2}}{4} \m \nm , \\
	\nu_2 &=& - \frac{\sqrt{2}}{4}\mm .    
	\eeqa  
We have identified $\nu_1 $ and $\nu_2$ in the comparison. 
Both of the periods are now ``quantized'' by the winding numbers 
$n_i$ and $n_{i}'$. This fact is characteristic to massive theories.    

To end this section, let us comment on the monodromy. 
From (\ref{a2}) and (\ref{aD2}), we can easily find that 
the monodromy matrix $M_{2,\infty}$ around $u=\infty$ acts to the 
column vector $v_2 = \ ^t (a_{D}^2 ,a_2 ,\epsilon_2 )$ as $v_2 \longrightarrow 
M_{2,\infty} \cdot v_2 $, i.e, 
	\beq
	\left(
	\begin{array}{c}
	 a_{D}^2 \\
	a_2 \\
	\epsilon_2 
	\end{array}\right) \longrightarrow 
	\left(\begin{array}{rrc}
	-1&2 & \mu_2  \\
	0&-1 &2 \\
	0&0 &1 
	\end{array}\right)\left(
	\begin{array}{c}
	 a_{D}^2 \\
	a_2 \\
	\epsilon_2
	\end{array}\right), \lab{mono}
	\eeq
where $\epsilon_2 =n_1 \nu_1 +n_2 \nu_2$ and $\mu_2 =
2(n_{1} ' -n_1 )/n_1 =2(n_{2} ' -n_2 )/n_2 $. Note that in the mass-less limit 
we can easily recover the monodromy of the mass-less $N_f =2$ theory \cite{IY}. 

Finally, since the residue contributions can 
appear also in the periods near the regular singular points, the monodromy matrices 
near them will be quantized as well. Of course mass-less 
theories do not have such residue contributions, so we can conclude that the 
monodromy matrix can be quantized, in general, only in massive theories. 

\begin{center}
\section{Prepotential}
\end{center}
\renewcommand{\theequation}{3.\arabic{equation}}\setcounter{equation}{0}

We can obtain the prepotential ${\F}_2$ from the relation 
	\beq
	a_{D}^2   = \frac{d{\F}_2}{da_2}
	.\eeq 
For that purpose, we should express $a_{D}^2 $ as a series of $a_2$. 
However, since $a_2 $ has constant terms, 
it is convenient to use $\tilde{a}_2 =a_2 -n_1 \nu_1 -n_2 
\nu_2 $  as a new variable. Then $a_{D}^2 $ will be expanded as 
	\beqa
	a_{D}^2 &=& n_{1} '\nu_1 +n_{2} ' \nu_2 +\sqrt{2} \tilde{a}_2 (
	B-A\ln 2 -2A\ln \tilde{a}_2 ) + \frac{A}{\sqrt{2}} \left\{ \frac{1}{2\tilde{a}_2}(\m^2 +\mm^2)\right.\nm \\
	& & -\frac{1}{6144\tilde{a}_{2}^3}\left[3\lla^4 +384\lla^2 \m \mm -256(\m^4 +\mm^4)\right] \nm \\
	& & +\left. \frac{1}{30720\tilde{a}_{2}^5}\left[45\lla^4 (\m^2 +\mm^2 )+ 256(\m^6 +\mm^6 )\right] +\cdots 
	\right\} .\lab{4.2}
	\eeqa
Therefore the prepotential will be  
	\beqa
	{\F}_2 &=&i\frac{\tilde{a}_{2}^2}{\pi} \left[ \frac{1}{2}\ln 
	\left( \frac{\tilde{a}_2}{\lla}\right)^2  
	+\left( -1+\frac{i \pi}{2}+\frac{5}{2}\ln 2 \right) 
	-\frac{\sqrt{2}\pi}{4i\tilde{a}_2}(n_{1} ' \m +n_{2} '\mm)\right. 
	\nm \\
	& & -\left. \frac{ \ln \tilde{a}_2}{4\tilde{a}_{2}^2}(\m^2 +\mm^2 )  
	+\sum_{i=2}^{\infty} {\F}_{i}^2 \tilde{a}_{2}^{-2i} \right] , \lab{pre}
	\eeqa
where the first few coefficients of the prepotential are given by  
	\beqa
	{\F}_{2}^2 &=& -\frac{\lla^4}{8192} +\frac{1}{96}(\m^4 +\mm^4 )
	-\frac{\lla^2 }{64}\m\mm , \nm \\
	{\F}_{3}^2 &=& \frac{3\lla^4 }{16384}(\m^2 +\mm^2 ) +
	\frac{1}{960}(\m^6 +\mm^6 ) ,\nm \\
	{\F}_{4}^2 &=& -\frac{5\lla^8}{268435456} +\frac{1}{5376}(\m^8 +\mm^8)  
	-\frac{5\lla^6 }{393216}\m\mm -\frac{5\lla^4 }{32768}\m^2 \mm^2 . \lab{resul}
	\eeqa
We find that these expansion coefficients have the same structure as $a_{2,n}$ or 
$b_{2,n}$. In the mass-less limit, we can easily recover the result of Ref.9. 
	
The reader may notice that this massive prepotential contains a curious term proportional to 
$(\m^2 +\mm^2 )\ln \tilde{a}_2$, but we can observe that 
	\beqa
	{\F}_{s}^{2}&= &\sum_{i=1}^{2}\left( \tilde{a}_2 -\frac{m_i}{\sqrt{2}}\right)^2 
	\ln \left(\tilde{a}_2 -\frac{m_i}{\sqrt{2}}\right) 
	+\sum_{i=1}^{2}\left( \tilde{a}_2 +\frac{m_i}{\sqrt{2}} \right)^2 
	\ln \left(\tilde{a}_2 +\frac{m_i}{\sqrt{2}}\right)\nm \\
	&= &(\m^2+\mm^2) \left(\ln \tilde{a}_2 +\frac{3}{2}\right) +4\tilde{a}_{2}^2 
	\ln \tilde{a}_{2} -\frac{\m^4 +\mm^4}{24\tilde{a}_{2}^2}-\frac{\m^6 +\mm^6}
	{240\tilde{a}_{2}^4}+\cdots .
	\eeqa
Therefore we can rewrite (\ref{pre}) as 
 	\beqa
	{\F}_2 &=&i\frac{\tilde{a}_{2}^2}{\pi} \left[ \frac{1}{2}\ln 
	\left( \frac{\tilde{a}_2}{\lla}\right)^2  
	+\left( -1+\frac{i \pi}{2}+\frac{5}{2}\ln 2 \right) 
	-\frac{\sqrt{2}\pi}{4i\tilde{a}_2}(n_{1} ' \m +n_{2} '\mm) \right. 
	\nm \\
	& &+\ln \tilde{a}_2 +\left. \frac{3}{8\tilde{a}_{2}^2}(\m^2 +\mm^2 ) -\frac{1}{4\tilde{a}_{2}^2}
	{\F}_{s}^2 +
	\sum_{i=2}^{\infty}
	{\widetilde{\F}}_{i}^2 \tilde{a}_{2}^{-2i}  \right],
	\eeqa	
where 
	\beqa
	\widetilde{\F}_{2}^2 &=& -\frac{\lla^4}{8192}-\frac{\lla^2 }{64}\m\mm , \nm \\
	\widetilde{\F}_{3}^2 &=& \frac{3\lla^4 }{16384}(\m^2 +\mm^2 ) ,\nm \\
	\widetilde{\F}_{4}^2 &=& -\frac{5\lla^8}{268435456} -\frac{5\lla^6 }{393216}\m\mm 
	-\frac{5\lla^4 }{32768}\m^2 \mm^2 . 
	\eeqa

\section{\protect\centering Double scaling limit}
\renewcommand{\theequation}{4.\arabic{equation}}\setcounter{equation}{0}

In this section, we discuss the double scaling limit $(\mm 
\longrightarrow \infty, 
\lla \longrightarrow 0, \mm \lla^2 =\la^3 \ \mbox{fixed})$ 
of the massive $N_f =2$ theory. We may scale $\m$ instead of $\mm$, holding 
$\m \lla^2$ fixed. Since the low energy theory of the massive 
$N_f =2$ can be regarded as the massive $N_f =1$ theory\cite{SW2}, 
we can check the consistency of our calculation by the double scaling 
limit. Discussions on the double scaling limit for the massive $N_f =1$ 
theory, i.e, reduction from the massive $N_f =1$ to the $N_f =0$ theory 
can be found in Ref.10. 

First, let us discuss the Picard-Fuchs equation (\ref{pic}). 
In the double scaling limit, coefficients (\ref{coepic}) will be 
	\beqa
	& & A_{2} \longrightarrow A_{2\mbox{\scriptsize dsl}}=64 \mm^2 \cdot (4u^2 -3\m \la^3 ), \nm \\
	& & B_{2} \longrightarrow B_{2\mbox{\scriptsize dsl}} =64\mm^2 \cdot (12u-16\m^2 ),\nm \\
	& & C_{2} \longrightarrow C_{2\mbox{\scriptsize dsl}}= 32\mm^2 \cdot  (3u-2\m^2 ), \nm \\
	& & D_{2} \longrightarrow D_{2\mbox{\scriptsize dsl}}= 32 \mm^2 \cdot 3,\nm \\
	& & \Del_{2} \longrightarrow \Del_{2\mbox{\scriptsize dsl}}=
	-16\mm^2 \cdot \left[27\la^6 -32\m \la^3 (9u-8\m^2 )-256(\m^2 u^2 -u^3 )\right].
	\eeqa	
We can rewrite these coefficients to more convenient forms  
	\beqa 
	& &A_{2 \mbox{\scriptsize dsl}} =-64 \mm^2 A_1 ,\ B_{2\mbox{\scriptsize dsl}} =-64\mm^2 B_1,
	\ C_{2 \mbox{\scriptsize dsl}} =-32\mm^2 C_1,\nm \\ 
	& &D_{2 \mbox{\scriptsize dsl}} =-32 \mm^2 D_1 ,\ 
	\Del_{2 \mbox{\scriptsize dsl}} =-16 \mm^2 \Del_1 ,\lab{redcoepic}
	\eeqa
where $A_1$, etc are the coefficients of the massive $N_f =1$ Picard-Fuchs equation. See appendix 
D. It is interesting to note that $A_{2 \mbox{\scriptsize dsl}}$, etc are $\mm^2$ multiples of 
$A_1$, etc , respectively. From (\ref{redcoepic}), we find that 
(\ref{pic}) in the double scaling limit reduces to 
	\beq
	\frac{d^3 \Pi_2}{du^3} +\left( \frac{\Del_{1} '}{\Del_1} 
	-4\frac{D_1}{B_1}
	\right) \frac{d^2 \Pi_2}{du^2}-\frac{32}{\Del_1} 
	\left( C_1 -\frac{A_1 D_1}{B_1} \right) 
	\frac{d\Pi_2 }{du} =0 \lab{redpic}.
	\eeq
This looks the Picard-Fuchs equation for the massive $N_f =1$ theory \cite{O}. 
However, since it is unclear whether $\Pi_2$ always reduces to $\Pi_1$, the period integral of the 
massive $N_f =1$ theory, there is no assurance that the solution $\Pi_2$ to (\ref{redpic}) 
equals to $\Pi_1$ exactly. 	

In fact, $\lambda_2$ in the double scaling limit will behave as 
	\beqa
	\lambda_2 &\longrightarrow& \frac{\sqrt{2}xdx}{4\pi i \tilde{y}}
	\left[ \frac{x^2 -u}{2} \left(\frac{1}{x+\m}+\frac{1}{x+\mm}\right)  -2x \right] \nm \\
	&=& \frac{\sqrt{2}xdx}{4\pi i \tilde{y}} \left[\frac{x^2 -u}{2(x+\m)}
	-2x +\frac{x^2 -u}{2\mm}\left( 1-\frac{x}{\mm}+\cdots \right) \right] \nm \\
	&=& \lambda_1 + \frac{\sqrt{2}xdx}{4\pi i \tilde{y}} \cdot 
	\frac{x^2 -u}{2\mm}\left(1-\frac{x}{\mm}+\cdots \right),
	\eeqa
where $\tilde{y}^2 =(x^2 -u)^2 -\la^3 (x+\m)$ is the curve for the massive 
$N_f =1$ theory and $\lambda_1$ is its meromorphic 1-form. 
We find that $\lambda_2$ in the double scaling limit consists 
of $\lambda_1$ and an extra 1-form. Of course, this extra 1-form under $\mm 
\longrightarrow 
\infty$ vanishes, but we can expect that divergences related to large $\mm$ should 
appear in our solutions because we have calculated all quantities under the premise 
that $\lla,\m$ and $\mm$ are finite. 
However we can drop the divergences because the heavy quark can be 
integrated out in taking the double scaling limit \cite{SW2}. Then, $\Pi_2$ in 
(\ref{redpic}) will be equal to $\Pi_1$ up to irrelevant divergences.   	

For example, for the residue contributions, we can 
eliminate ``$\nu_2 $'' which depends on $\mm$ while keeping $\nu_1$. 
We can show that $a_2$ in the double scaling limit coincides with 
$a_1 $, the corresponding period in the massive $N_f =1$ theory. This fact 
means that $a_2$ does not receive any effects originated from the extra 1-form. 
On the other hand, the effects are non-trivial for $a_{D}^2 $. In order to 
extract them, it is better 
to rearrange it by $a_2 $. Namely, we will be able to
 divide $a_{D}^2 $ into the convergent part and the divergent one. For that purpose, it is convenient 
to use (\ref{4.2}). Then we can actually see that the expansion coefficients in the double scaling limit 
are a sum of the finite part and the divergent one. Since this divergence depends only on $\mm$, 
we can eliminate it. In this case, $a_{D}^2 $ coincides with the corresponding 
period $a_{D}^1$ of the massive $N_f =1$ theory. 
In fact, we can easily find that we can obtain $a_{D}^1$ if we drop the $\mm$ dependences of (\ref{4.2}) 
and rewrite it as a series of $u$, but the constants $A$ and $B$ which 
correspond to the initial conditions for the Picard-Fuchs equation should be replaced 
with those of the massive $N_f =1$ theory \cite{O}. 

Dropping these divergences and integrating $a_{D}^2 $ over ``$a_1$'' 
to obtain the prepotential in the double scaling limit, we find that 
the instanton expansion coefficients are   
	\beqa
	\widetilde{\F}_{2}^2 &\longrightarrow& -\frac{1}{64}\la^3 \m ,
	\nm \\
	\widetilde{\F}_{3}^2 &\longrightarrow& \frac{3\la^6}{16384}, 
	\nm \\
	\widetilde{\F}_{4}^2 &\longrightarrow& -\frac{5\la^6 \m^2}
	{32768} .
	\eeqa
These are nothing other than the instanton 
expansion coefficients of the massive $N_f =1$ theory \cite{O}! 
The other asymptotic leading terms coincide with those of the massive $N_f =1$ after the 
replacement of $A$ and $B$, we do not write down them here. These can be easily obtained from (\ref{form}) 
with $N_f =1$ (see below). In this way, we can obtain the prepotential of the massive $N_f =1$ theory.

\begin{center}
\section{Summary}
\end{center}
\renewcommand{\theequation}{5.\arabic{equation}}\setcounter{equation}{0}

We have studied the moduli space of $N=2$ $SU(2)$ Yang-Mills gauge theory 
coupled with $N_f =2,3$ massive matter multiplets and 
clarified the relation between the massive theories by double scaling limit.
Though we have mainly stated on the massive $N_f =2$ theory in the previous sections, 
by using the results in appendix C, Refs.5 and 10, 
we can write the periods of the massive $N_f$ $(N_f =0,\cdots ,3)$ theory as 
	\beqa
	a_{N_f}(u)& =&  -\frac{\sqrt{2}}{4}\sum_{i=1}^{N_f}n_i  m_i  
	+\frac{1}{2}\sqrt{2u}\left[ 1+\sum_{i=1}^{\infty} a_{N_f ,i} 
	(\Lambda_{N_f}^{4-N_f},\m,\cdots,m_{N_f}) u^{-i} \right] ,\nm \\
	a_{D}^{N_f} (u)& =& -\frac{\sqrt{2}}{4}\sum_{i=1}^{N_f} n_{i} ' m_i +
	i\frac{4-N_f}{2\pi}\tilde{a}_{N_f}(u) 
	\ln \left(\frac{u}{\Lambda_{N_f}^2}
	\right) \nm \\	
	& & +\sqrt{u} \sum_{i=0}^{\infty} a_{D_i} (\Lambda_{N_f}^{4-N_f} 
	,\m,\cdots,m_{N_f} ) u^{-i}, \lab{conj} 
	\eeqa
where $a_{N_f ,i} (\Lambda_{N_f}^{4-N_f},\m,\cdots,m_{N_f})$ and $a_{D_{i}} 
(\Lambda_{N_f}^{4-N_f},\m,\cdots,m_{N_f})$ are homogeneous polynomials of order $2i$ and are invariant 
under the obvious $\mbox{\boldmath$Z$}_{N_f}$ symmetry. 
Then, from (\ref{conj}) we can easily find that the monodromy matrices 
$M_{N_f ,\infty}$'s around $u=\infty$ act to the column vectors 
$v_{N_f} =\ ^t ( a_{D}^{N_f}, a_{N_f},\epsilon_{N_f} )$ 
as $v_{N_f} \longrightarrow M_{N_f ,\infty}\cdot \nu_{N_f}$ and are given by 
	\beq
	M_{N_f ,\infty} =\left(\begin{array}{rcc}
	-1& 4-N_f & \mu_{N_f} \\
	0&-1  & 2   \\
	0&0&1 
	\end{array}\right), \lab{monodromy}
	\eeq
where $\epsilon_{N_f} =n_1 \nu_1  +\cdots +n_{N_f} \nu_{N_f}$, $\mu_{N_f} =2n_{N_f} ' /n_{N_f}
-(4-N_f ) $. The non-trivial 
relations among the winding numbers are 
	\beqa
	& &N_f =1   \ \ \ \ \ \ \ \ \ \ \ \ \ \ \ \ \ \ \stackrel{\ }{------} \nm \\
	& &N_f =2 \ \ \ \ \ \ \ \ \ \ \ \ \ \ \ \ \ \ \ \ n_{1} n_{2} ' =n_2 n_{1} '  , \nm \\
	& &N_f =3 \ \ \ \ \ \ \ \ \ \ n_1 n_2 n_{3}' =n_1 n_{2} ' n_3 =n_{1} ' n_2 n_3  
	.\eeqa
As for the prepotentials ${\F}_{N_f}$, we have established that they are 
given by the following simple formulae,
	\beqa
	{\F}_{N_f}& =& i\frac{\tilde{a}_{N_f}^2}{\pi}\left[ \frac{4-N_f}{4}\ln 
	\left( \frac{\tilde{a}_{N_f}}{\Lambda_{N_f}}\right)^2 +{\F}_{0}^{N_f} -
	\frac{\sqrt{2}\pi}{4i\tilde{a}_{N_f}}\sum_{i=1}^{N_f} n_{i} ' m_i \right.
	+\frac{N_f}{2}\ln \tilde{a}_{N_f}  \nm \\
	& &+\left. \frac{1}{4\tilde{a}_{N_f}^2}\left(
	\frac{3}{2} \sum_{i=1}^{N_f}m_{i}^2 -{\F}_{s}^{N_f} \right)+
	\sum_{i=2}^{\infty} \widetilde{\F}_{i}^{N_f} (\Lambda_{N_f}^{4-N_f} 
	,\m ,\cdots,m_{N_f}) 
	\tilde{a}_{N_f}^{-2i}   \right] \lab{form} ,
	\eeqa
where 
	\beq
	{\F}_{s}^{N_f} =\sum_{i=1}^{N_f}\left(\tilde{a}_{N_f}-\frac{m_i}{\sqrt{2}}\right)^2 \ln 
	\left(\tilde{a}_{N_f}-\frac{m_i}{\sqrt{2}}\right) +\sum_{i=1}^{N_f}\left(\tilde{a}_{N_f}+\frac{m_i}{\sqrt{2}}\right)^2 \ln 
	\left(\tilde{a}_{N_f}+\frac{m_i}{\sqrt{2}} \right)
	,\eeq
and ${\F}_{0}^{N_f}$'s are some calculable constants 
independent of $\Lambda_{N_f} $ and $m_i$. Here we have used 
$\tilde{a}_{N_f} =a_{N_f} - (n_1 \nu_1 +\cdots +n_{N_f} \nu_{N_f})$. Note that 
$\widetilde{\F}_{i}^{N_f}$ 's are homogeneous polynomials.

Finally, we comment on some open problems. 
Though we 
have restricted ourselves within the  discussions in the weak coupling limit, 
it is important to quantitatively investigate in the strong coupling region. However, 
since there are many technical obstacles to accomplish it in the massive theories, 
it may be useful to do it by using some integrable systems \cite{EY,IM}. 
If the investigations in the strong coupling region are done, 
we will be able to sufficiently understand the properties of 
these massive asymptotic free gauge theories. 
In addition to this, we should also 
check whether these formulae can be obtained by some 
field theoretical method.

\section*{\protect\centering Acknowledgement}

The author acknowledge interesting discussions with Dr. H. Kanno. 

\begin{center}
\section*{Appendix A Expansion coefficients (I) }
\end{center}
\renewcommand{\theequation}{A\arabic{equation}}\setcounter{equation}{0}

The first several coefficients of $\rho_1$ are
	\beqa
	a_{2,0} &=& 1 \nm ,\\
	a_{2,1} &=&0 \nm, \\
	a_{2,2} &=& -\frac{1}{1024}(\lla^4 +64\lla^2 \m \mm),\nm \\
	a_{2,3} &=& \frac{3\lla^4}{1024}(\m^2 +\mm^2) ,\nm \\
	a_{2,4} &=& -\frac{15\lla^4}{4194304}(\lla^4 +256\lla^2 \m \mm +4096\m^2 \mm^2) ,\nm \\
	a_{2,5} &=& \frac{35\lla^6}{2097152}(\m^2 +\mm^2 )(3\lla^2 +128\m \mm),
	 \nm \\
	a_{2,6} & =& -\frac{105\lla^6}{4294967296}\left [\lla^6 +3072\lla^2 (\m^4 +\mm^4 )+576\lla^4 \m \mm 
	+36864\lla^2 \m^2 \mm^2 \right. \nm \\
	& &\left. +262144\m^3 \mm^3  \right].
	\eeqa

\section*{\protect\centering Appendix B Expansion coefficients (II) }
\renewcommand{\theequation}{B\arabic{equation}}\setcounter{equation}{0}

The first several coefficients of $\rho_2$ are
	\beqa
	b_{2,1} &=& \frac{1}{2}(\m^2 +\mm^2 ) \nm ,\\
	b_{2,2} &=& \frac{1}{3072}\left[3\lla^4 +256(\m^4 +\mm^4 )\right],\nm \\
	b_{2,3} &=& \frac{\m^2 +\mm^2 }{30720}\left[15\lla^4  +960\lla^2 \m \mm +1024(\m^4-\m^2 \mm^2 +\mm^4 )\right] \nm ,\\	
	b_{2,4} &=& \frac{1}{58720256}\left[91 \lla^8 - 71680 \lla^4 (\m^4 +\mm^4 ) + 1048576(\m^8 +\mm^8 )- 3584 \lla^6 \m \mm 
	\right.\nm \\
	& & \left. +917504 \lla^2 (\m^5 \mm + \m \mm^5 -  \lla^2 \m^2 \mm^2 )\right],  \nm \\
	b_{2,5} &=&\frac{\m^2 + \mm^2}{377487360}\left[ 1935 \lla^8 - 215040 \lla^4 (\m^4 + \mm^4 )+ 4194304(\m^8 +\mm^8 )\right.
	\nm \\
	& &+ 687360 \lla^6 \m \mm  + 3440640 \lla^4 \m^2 \mm^2 + 4194304(\m^4 \mm^4 - \m^2 \mm^6 - \m^6 \mm^2 )\nm \\
	& &\left. + 3932160 \lla^2 (\m \mm^5 +\m^5 \mm -  \m^3 \mm^3 ) \right] ,\nm \\
	b_{2,6} &=&\frac{1}{283467841536}\left[1793 \lla^{12} - 17554944 \lla^8 (\m^4 + \mm^4 )-103809024\lla^4 (\m^8 +\mm^8 ) \right. \nm \\
	& & - 297792 \lla^{10} \m \mm  +2214592512 \lla^2 (\m^9 \mm +\m \mm^9 ) - 146792448 \lla^8 \m^2 \mm^2 \nm \\
	& &+1453326336 \lla^4 (\m^6 \mm^2 +\m^2 \mm^6 ) -2860515328 \lla^6 \m^3 \mm^3 \nm \\
	& & \left.- 272498688 \lla^6 (\m \mm^5 +\m^5 \mm) + 2147483648(\mm^{12}+\m^{12} ) \right].
	\eeqa

\begin{center}
\section*{Appendix C Results of the massive $N_f =3$ theory }
\end{center}
\renewcommand{\theequation}{C\arabic{equation}}\setcounter{equation}{0}

In this appendix, we summarize the results for the massive $N_f =3$ theory. 
We use the same notations as in the text, unless we mention especially. 

In this theory, its quantum moduli space can be described by 
the following hyperelliptic curve 
	\beq
	y^2 =F(x)^2 -G(x), \lab{N3curve}
	\eeq
and the meromorphic 1-form
	\beq
	\lambda_3 = \frac{\sqrt{2}xdx}{4\pi i y} \left[ \frac{F(x) G'(x)}{2G(x)} -F'(x)
	\right]
	,\eeq
where 
	\beqa
	F(x)&=& x^2 -u +\Lambda_3 \left( \frac{\m+\mm+m_3 }{8} +\frac{x}{4}\right) , \nm \\
	G(x)&=& \Lambda_3 (x+\m) (x+\mm) (x+m_3) .
	\eeqa
The prime denotes the differentiation over $x$. 
Four branching points of (\ref{N3curve}) in the weak coupling limit are given by
	\beqa
	x_1 &=&\frac{\llla}{8}+\sqrt{u}+\frac{\sqrt{\llla}}{2}u^{1/4} +\frac{\sqrt{\llla}}{4u^{1/4}}
	\left(\m+\mm+\mmm+\frac{\llla}{16}\right) \nm \\
	& &+\frac{\llla}{16\sqrt{u}}(\m+\mm+\mmm) +\cdots ,\nm \\
	x_2 &=& \frac{\llla}{8}+\sqrt{u}-\frac{\sqrt{\llla}}{2}u^{1/4} -\frac{\sqrt{\llla}}{4u^{1/4}}
	\left(\m+\mm+\mmm+\frac{\llla}{16}\right) \nm \\
	& &+\frac{\llla}{16\sqrt{u}}(\m+\mm+\mmm) +\cdots ,\nm \\
	x_3 &=& \frac{\llla}{8}-\sqrt{u}+i\frac{\sqrt{\llla}}{2}u^{1/4} -\frac{i\sqrt{\llla}}{4u^{1/4}}
	\left(\m+\mm+\mmm+\frac{\llla}{16}\right) \nm \\
	& &-\frac{\llla}{16\sqrt{u}}(\m+\mm+\mmm) +\cdots ,\nm \\
	x_4 &=& \frac{\llla}{8}-\sqrt{u}-i\frac{\sqrt{\llla}}{2}u^{1/4} +\frac{i\sqrt{\llla}}{4u^{1/4}}
	\left(\m+\mm+\mmm+\frac{\llla}{16}\right) \nm \\
	& &-\frac{\llla}{16\sqrt{u}}(\m+\mm+\mmm) +\cdots .
	\eeqa
Then the two periods of $\lambda_3$ are defined by 
	\beqa
	a_3 (u)&=& \oint_{\alpha'} \lambda_3 \nm, \\
	a_{D}^3 (u) &=& \oint_{\beta'} \lambda_3 ,
	\eeqa
where $\alpha$ and $\beta$ are loops which may enclose the ``extra'' poles corresponding to 
$x=-\m,-\mm, -m_3$. We can identify the canonical $\alpha$-cycle with a loop which enclose the 
two branching points from $x_4$ to $x_1$ counter clockwise and the canonical $\beta$-cycle with a 
loop from $x_1 $ to $x_2$ as well.  

Using $\Pi_3 =\oint_{\gamma} \lambda_3$, where $\gamma$ is a suitable 1-cycle on the curve, we can 
obtain the massive $N_f =3$ Picard-Fuchs equation 
	\beq
	\frac{d^3 \Pi_3}{du^3}+\left(\frac{\Del_{3}'}{\Del_3 }
	-\frac{4D_3}{B_3}\right)\frac{d^2\Pi_3}{du^2}
	-\frac{256}{\Del_3 }\left(C_3  -\frac{A_3 D_3}{B_3}\right)\frac{d\Pi_3}{du} =0, \lab{pic3}
	\eeq
where
	\beqa
	A_3 &=& \llla^4 K_1 (K_3 -L_1 +S_1 )+2048u^2 (3u^2 +N_2 -2u K_2 ) \nm \\
	& & +4 \llla^3 [-2K_{5} -10S_1 K_2 +2S_1 N_1 +2L_3 -2H_4 +u ( K_3 +L_1 +18S_1 ) ] \nm \\
	& & +256\llla \left[ -u^3 K_1 -6S_1 N_2 +4S_2 K_1 \right. \nm \\
	& &\left. +2u^2 (K_3 +L_1 -12S_1 )-u( 3L_3 +3S_1 N_1  -14S_1 K_{2} )\right]  \nm \\
	& & +32\llla^2 \left[ -5u^2 K_2 +u(2K_4 +18N_2 -6S_1 K_1)-6L_4 +2S_1 K_3 +2S_1 L_1 -6S_2 \right] ,\nm \\
	B_3 &=& 4K_2 (\llla^2 u+16\llla S_1 +128u^2 )+8N_2 (\llla^2 -96u) -S_1 (\llla^3 +192\llla u) \nm \\
	& & -256(u^3 -4S_2 )-8\llla^2 K_4 ,\nm \\
	C_3 &=& \llla^3 (K_3 +L_1 +16 S_1 )+8\llla^2 (2K_4 -6K_1 S_1 +16 N_2 -9uK_2 ) \nm \\
	& &+512(11u^3 -2S_2 -6u^2 K_2 +3u N_2 ) \nm \\
	& & +64\llla \left[ -3 L_3 -3u^2 K_1 +12S_1 K_2 +u(4K_3 +4L_1 -42S_1 ) -3S_1 N_1 \right], \nm \\
	D_3 &=&K_2 (\llla^2 +256u) -48\llla S_1 -192 (u^2 +N_2) ,\nm \\
	\Del_3 &=& -1048576(u^5 +\llla S_3)-32\llla^5 S_1 K_2 +4096u^2 (\llla^2 +256K_2) \nm \\
	& &+\llla^6 (K_4 -2N_2 )-6144\llla^3 S_1 (2K_4 +N_2) -12288\llla^2 (9N_4 +2S_2 K_2 ) \nm \\
	& &+ 128\llla^4 (2K_6 -13S_2 -3L_4 ) +u^3 \left[ 32768\llla^2 K_2 +131072(11\llla S_1 -8N_2)\right] \nm \\
	& &+u^2 \left[ 2048S_1 (-11\llla^3 +512S_1)-128\llla^4 K_2 -1310720\llla S_1 K_2 \right. \nm \\
	& &\left.-8192\llla^2 (4K_4 +23N_2 )\right] +u\left[ 64\llla^5 S_1 +26624\llla^3 S_1 K_2 \right.\nm \\	
	& &+\left. 1179648\llla S_1 N_2 +256\llla^4 (4N_2 -K_4 ) +16384\llla^2 (5S_2 +9L_4 )\right] ,\nm \\
	S_i &=& (\m\mm\mmm)^i ,\nm \\
	K_i &=& \m^i +\mm^i +\mmm^i ,\nm \\
	L_i &=& \m^i (\mm^2 +\mmm^2 ) +\mm^i (\mmm^2 +\m^2 )+\mmm^i (\m^2 +\mm^2 ) ,\nm \\
	N_i &=& (\m\mm)^i  +(\mm \mmm)^i +(\mmm\m)^i  ,\nm \\
	H_4 &=& \m^4 (\mm+\mmm) +\mm^4 (\mmm +\m)+\mmm^4 (\m+\mm) .
	\eeqa
In the double scaling limit $(\llla \longrightarrow 0, \mmm \longrightarrow \infty, \llla \mmm =\lla^2 \ \mbox{fixed})$, 
they will be 
	\beqa
	A_3&\longrightarrow &A_{3\mbox{\scriptsize dsl}} = 8 \mmm^2 A_2 , \nm \\ 
	B_3 &\longrightarrow & B_{3\mbox{\scriptsize dsl}} =-\mmm^2 B_2 ,\nm \\ 
	C_3 &\longrightarrow & C_{3\mbox{\scriptsize dsl}} =16 \mmm^2 C_2 ,\nm \\
	D_3 &\longrightarrow & D_{3\mbox{\scriptsize dsl}}=-2\mmm^2 D_2  , \nm \\
	\Del_3 &\longrightarrow &\Del_{3\mbox{\scriptsize dsl}} =256 \mmm^2 \Del_2 .
	\eeqa
Using these coefficients, we can obtain (\ref{pic}) as a 
resultant of the double scaling limit. 

The fundamental solutions to (\ref{pic3}) near $u=\infty$ $(z=1/u)$ are given by
	\beqa
	\tilde{\rho}_0 (z)&=& -\frac{\sqrt{2}}{4} \sum_{i=1}^{3} n_i m_i ,\nm \\
	\tilde{\rho}_1 (z)&=& z^{-1/2} \sum_{i=0}^{\infty} a_{3,i} z^i ,\nm \\
	\tilde{\rho}_2 (z) &=& \tilde{\rho}_1 (z) \ln z +z^{-1/2} \sum_{i=1}^{\infty} b_{3,i} z^i ,
	\eeqa
where
	\beqa
	a_{3,0} &=& 1 , \nm \\
	a_{3,1} &=& -\frac{\llla^2}{1024} , \nm \\
	a_{3,2} &=& -\frac{3\llla^4}{4194304}-\frac{\llla^2}{1024}(\m^2+\mm^2+\mmm^2)- \frac{\llla}{16}
		\m\mm\mmm \nm , \\
	a_{3,3} &=& -\frac{5\llla^6}{4294967296}-\frac{3\llla^4}{2097152}(\m^2 +\mm^2 +\mmm^2)+\frac{3\llla^3}{16384}\m\mm\mmm
		 \nm \\
		& & +\frac{3\llla^2}{1024}(\m^2\mm^2 +\mm^2 \mmm^2 +\mmm^2 \m^2) ,\nm \\
	a_{3,4} &=& -\frac{175\llla^8}{70368744177664}-\frac{15\llla^6}{4294967296}(\m^2+\mm^2+\mmm^2)+\frac{15\llla^5 }{67108864}\m\mm\mmm \nm \\
		& &- \frac{15\llla^4}{1048576}\left[\frac{1}{4}(\m^4+\mm^4+\mmm^4) -(\m^2\mm^2+\mm^2\mmm^2 +\mmm^2\m^2)\right] \nm \\
		& &-\frac{15\llla^3}{16384}\m\mm\mmm(\m^2+\mm^2+\mmm^2) -\frac{15\llla^2}{1024}\m^2 \mm^2 \mmm^2 
	\eeqa
and
	\beqa
	b_{3,1} &=& \m^2 +\mm^2+\mmm^2 +\frac{\llla^2}{512} ,\nm \\
	b_{3,2} &=& \frac{\llla^4}{4194304}+\frac{\llla^2 }{1024}(\m^2+\mm^2+\mmm^2) -\frac{\llla}{8}\m\mm\mmm 
		+\frac{1}{6}(\m^4+\mm^4+\mmm^4) ,\nm \\
	b_{3,3} &=& -\frac{\llla^6}{12884901888}-\frac{\llla^4}{4194304}(\m^2+\mm^2+\mmm^2) -\frac{\llla^3}{4096}\m\mm\mmm\nm \\
		& &+\frac{\llla^2}{128} \left[\frac{3}{16}(\m^4+\mm^4+\mmm^4) +\m^2\mm^2+\mm^2\mmm^2+\mmm^2\m^2 \right] \nm \\
		& & +\frac{\llla}{16}\m\mm\mmm(\m^2 +\mm^2 +\mmm^2 ) +\frac{1}{15}(\m^6+\mm^6+\mmm^6) ,\nm \\
	b_{3,4} &=& -\frac{265\llla^8}{422212465065984}-\frac{3\llla^6}{2147483648}(\m^2+\mm^2+\mmm^2) +\frac{\llla^5}{67108864}\m\mm\mmm \nm \\
		& &+\frac{\llla^4}{524288}\left[ \frac{13}{16}(\m^4+\mm^4+\mmm^4) +7(\m^2\mm^2 +\mm^2\mmm^2+\mmm^2\m^2) \right] \nm \\
		& & -\frac{\llla^2}{1024}\left[ \frac{5}{2}(\mm^2+\mmm^2)(\m^2\mm^2+\mm^2\mmm^2+\m^4)  
		+\frac{5}{6}(\m^6+\mm^6+\mmm^6) +61 \m^2\mm^2\mmm^2 \right] \nm \\
		& &-\frac{31\llla^3}{16384}\m\mm\mmm(\m^2 +\mm^2 +\mmm^2 )+\frac{\llla}{32}\m\mm\mmm(\m^4 +\mm^4 +\mmm^4 ) \nm \\
		& &+\frac{1}{28}(\m^8+\mm^8+\mmm^8) .
	\eeqa

From the lower oredr expansion of the periods, we find that 
	\beqa
	a_3 (u) &=& \frac{\tilde{\rho}_1 (z)}{\sqrt{2}} +\sum_{i=1}^{3} n_i \nu_i , \nm \\
	a_{D}^3 (u)&=& A' \tilde{\rho}_2 (z) +B' \tilde{\rho}_1 (z)+\sum_{i=1}^3 n_{i} ' \nu_i , 
	\eeqa
where
	\beqa
	A' &=& -\frac{i\sqrt{2}}{4\pi}, \nm \\
	B' &= &\frac{i\sqrt{2}}{4\pi}(8\ln 2 -1 -\pi i -2\ln \llla ),\nm \\
	\nu_i &=& -\frac{\sqrt{2}}{4} m_i 
	.\eeqa
Then the monodromy of the periods near $u=\infty$ will be 
	\beqa
	a_3 &\longrightarrow& -a_3 +2\sum_{i=1}^{3}n_i \nu_i , \nm \\
	a_{D}^3 &\longrightarrow& -a_{D}^3 +a_3 +\sum_{i=1}^3 (2n_{i} ' -n_i )\nu_i .
	\eeqa
 
Prepotential will be
	\beqa
	{\F}_3 &=& i\frac{\tilde{a}_{3}^2}{\pi}\left[ \frac{1}{4}\ln \left(\frac{\tilde{a}_3}{\llla}\right)^2 +\frac{1}{4}
	( 9\ln 2 -2 -\pi i ) -\frac{\sqrt{2}\pi}{4i\tilde{a_3}}\sum_{i=1}^{3} n_i ' m_i  \right. \nm \\  
	& &\left. -\frac{\ln \tilde{a}_3}{4}
	\sum_{i=1}^{3} m_{i}^2  + \sum_{i=2}^{\infty} {\F}_{i}^3 \tilde{a}_{3}^{-2i}\right],
	\eeqa
where first few instanton expansion coefficients are given by 
	\beqa
	{\F}_{2}^3 &=& -\frac{ \llla^4}{67108864}-\frac{\llla^2 }{8192}(\m^2 +\mm^2 +\mmm^2 )-\frac{\llla}{64}\m\mm\mmm 
	 +\frac{1}{96}(\m^4 +\mm^4 +\mmm^4 ) ,\nm \\
	{\F}_{3}^3 &=& \frac{3\llla^4}{67108864}(\m^2 +\mm^2 +\mmm^2 )+\frac{\llla^3}{65536} \m\mm\mmm \nm \\
	& &+\frac{3\llla^2}{16384}(\m^2\mm^2 +\mm^2\mmm^2 +\mmm^2\m^2)+\frac{1}{960}(\m^6+\mm^6+\mmm^6) ,\nm \\
	{\F}_{4}^3 &=&- \frac{5\llla^8}{9007199254740992} -\frac{5\llla^6}{206158430208}(\m^2+\mm^2+\mmm^2) 
	 -\frac{7\llla^5}{536870912} \m\mm\mmm \nm \\
	& &-\frac{5\llla^4 }{67108864} \left[ \frac{1}{4}(\m^4+\mm^4+\mmm^4 )+5(\m^2\mm^2+\mm^2\mmm^2+\mmm^2\m^2) \right] \nm \\
	& & -\frac{5\llla^3}{393216}\m\mm\mmm(\m^2+\mm^2+\mmm^2 ) -\frac{5}{32768}\llla^2\m^2\mm^2\mmm^2 \nm \\
	& &  +\frac{1}{5376}(\m^8+\mm^8+\mmm^8) \lab{N3ins} 
	.\eeqa
We 
can easily find that (\ref{N3ins}) coincide with those of Ref.9 when the three 
hypermultiplets are mass-less and that these coefficients does 
not vanish in general while ${\F}_{2n+1}^3 $ $(n>0)$ vanish in the mass-less limit. 
Note that we can rewrite these coefficients by using $\widetilde{\F}_{i}^3$, although 
we do not rewrite them here.

\section*{\protect\centering Appendix D $N_f =1$ Picard-Fuchs equation }
\renewcommand{\theequation}{D\arabic{equation}}\setcounter{equation}{0}

In this appendix, we briefly summarize 
the $N_f =1$ Picard-Fuchs equation. For more systematic explanations, see Ref.10. 
The curve and the meromorphic 1-form of the $N_f =1$ theory 
is given by 
	\beq
	y^2 =(x^2 -u)^2 -\la ^3 (x+\m),\lab{curve}
	\eeq
and 
	\beq
	\lambda_1 =\frac{\sqrt{2}xdx}
	{4\pi iy}\left[ \frac{x^2 -u}{2(x+\m)}-2x \right], 
	\eeq
respectively. 
Then the Picard-Fuchs equation is given by
	\beqa
	& &\frac{d^3 \Pi_1}{du^3} +\frac{3\DDel +\DDel' (4\m^2 -3u)}
	{\DDel (4\m^2 -3u)}
	\frac{d^2\Pi_1}{du^2}\nm \\
	&&-\frac{8[4(2\m^2-3u)(4\m^2-3u)+3(3\la^3 \m-4u^2 )]}{\DDel (4\m^2-3u)}
	\frac{d\Pi_1}{du} =0 , \lab{pic1}
	\eeqa
where 
	\beq
	\DDel =27\la^6 +256\la^3 \m^3 -288\la^3 \m u-256\m^2u^2+256u^3 ,
	\eeq
and $\DDel ' =d\DDel /du$. 
$\DDel $ is the discriminat of the curve. Introducing following notations 
	\beq
	A_1 =-4u^2 +3\m\la^3 ,\ B_1 =-12u+16 \m^2 ,\ 
	C_1 =-3u+2\m^2 ,\ D_1 =-3 ,
	\eeq
we can write (\ref{pic1}) more simply as 
	\beq
	\frac{d^3 \Pi_1}{du^3} +\left( \frac{\Del_{1} '}{\Del_1} 
	-\frac{4D_1}{B_1}
	\right) \frac{d^2 \Pi_1}{du^2}-\frac{32}{\Del_1} 
	\left( C_1 -\frac{A_1 D_1}{B_1} \right) 
	\frac{d\Pi_1 }{du} =0 ,
	\eeq
where the prime denotes the differentiation over $u$.

\begin{center}

\end{center}

\end{document}